\documentclass[11pt,a4paper]{article}

\usepackage{cite}
\usepackage{multicol,bbm}
\usepackage{amsmath, amssymb, amsfonts}
\usepackage{graphicx}
\usepackage{caption}
\usepackage{subcaption}
\usepackage{color}
\usepackage[utf8]{inputenc}
\usepackage{float}

\makeatletter \@addtoreset{equation}{section} \makeatother

\begin{document}

\title{Fermionic bound states in distinct kinklike backgrounds}

\author{D. Bazeia$^1$\thanks{bazeia@fisica.ufpb.br}\ \ and A. Mohammadi$^2$\thanks{a.mohammadi@fisica.ufpb.br}\\
\textit{$^1$Departamento de F\'{\i}sica, Universidade Federal da Para\'{\i}ba}\\\textit{58051-970 João Pessoa, Para\'{\i}ba, Brazil}\vspace{0.3cm}\\\textit{$^2$Departamento de F\'{\i}sica, Universidade Federal de Campina Grande}\\\textit{Caixa Postal 10071, 58429-900 Campina Grande, Para\'{\i}ba, Brazil}
}

\maketitle

\begin{abstract}
This work deals with fermions in the background of distinct localized structures in the two-dimensional spacetime. Although the structures have similar topological character, which is responsible for the appearance of fractionally charged excitations, we want to investigate how the geometric deformations that appear in the localized structures contribute to the change in the physical properties of the fermionic bound states. We investigate the two-kink and compact kinklike backgrounds, and consider two distinct boson-fermion interactions, one motivated by supersymmetry and the other described by the standard Yukawa coupling.
\end{abstract}

\maketitle

\section{Introduction}

The behavior of a fermion field in the background of a topological kink has been nicely studied in the pioneer work \cite{jr}. An important application of this investigation arises from the possibility that fermions in the presence of the kinklike structure can be physically realized in the polymeric chain known as polyacetylene \cite{ssh,ss,js}. It has also been suggested \cite{gw} that the fractionally charged states which appears in the topological background may support $1/2$ and other charge fractions.

In the work \cite{jr}, the background structure is the kink of the prototype Higgs field, represented by a scalar field $\phi$ with standard kinematics and a specific potential, that contains the fourth-order power in the field and develops spontaneous symmetry breaking. If one uses dimensionless units, the kinklike configuration engenders the standard form $\phi(x)=\tanh(x)$. In more recent years, however, distinct kinds of topological structures have been found in models described by a single real scalar field. Two examples of interest to the current work are the investigations carried out in Refs.~\cite{bmm,blmm}, in the first case in a model that led to the bosonic structures of the form $\phi(x)=\tanh^p(x)$, that produces an internal structure for $p$ being a positive odd integer greater or equal to $3$, and in the second, to a kinklike configuration controlled by the positive integer $n$, that shrinks the solution to a compact kinklike structure as $n$ increases to larger and larger values. Although the two structures have similar topological behavior, we think that it is of interest to investigate how the fermionic fields respond to these new geometrically distinct structures. 

The mathematical motivation is clear, but the issue is also of physical interest, since it has been shown experimentally in \cite{bischof} that under specific conditions, in constrained geometries the kink described by the standard $\tanh(x)$ configuration modifies to yield a profile similar to the new configuration $\tanh^p(x)$, for $p=3,5,7,...$; see Ref.~\cite{bmm}.
The investigation is of broader interest, since it can emerge in other contexts, in particular as planar topological structures in magnetic materials. The issue here is that the manipulation of vortices and skyrmions in miniaturized systems at the nanometric scale is now an experimental fact of practical importance; see, e.g., the works \cite{vor,nat,sci} and references therein. So, one should study the behavior of fermions not only in the new backgrounds that appears in \cite{bmm,blmm} in the case of one-dimensional structures, but also in the more general case of vortices and skyrmions in the case of two-dimensional structures. Another issue of current interest refers to the case studied in Ref. \cite{prl}, where one investigates the possibility to control the domain wall polarity by the action of current pulses. In particular, we notice that if the experiment is repeated in constrained geometries as the one investigated in \cite{bischof}, one can perhaps modify the way the domain wall polarity responds to the action of current pulses, since the geometry may change the internal structure of the configuration.

The purpose of the current work is to focus only on the behavior of fermions in the background of one-dimensional structures, such as the two-kink configuration described in \cite{bmm}, and the compact kinklike structure found in \cite{blmm}. We leave other issues and the study of fermions in the presence of two-dimensional structures to another work, to be described elsewhere. We organize the work in a way such that in Sec.~\ref{sec:gen} we review the basic facts on the subject, and in Sec.~\ref{sec:models} we study the new models considering two distinct couplings, highlighting the main differences among them. We end the work with comments and conclusions in Sec.~\ref{sec:end}.

\section{Generalities}
\label{sec:gen}

Let us start considering a boson field $\phi$ and a fermion field $\psi$ described by the minimal supersymmetric model defined in the $(1,1)$\,-\,dimensional spacetime. In this case, if one considers $W=W(\phi)$ as the superpotential, the supersymmetric Lagrangian density has the form \cite{Shifman:1998zy}
\begin{align}\label{a6}
	\mathcal{L} = \frac{1}{2}\left( \partial_\mu \phi\partial^\mu \phi +\bar{\psi}\ i\gamma^\mu\partial_\mu \psi + \mbox{F}^2 \right) + W_\phi\ \mbox{F}\ - \frac{1}{2}W_{\phi\phi}\bar{\psi}\psi,
\end{align}
where $\psi$ is a Dirac field with two independent degrees of freedom and $F$ is an auxiliary field. Also, the subscript $\phi$ stands for the derivative with respect to the scalar field. We can eliminate the field $F$ using its equation of motion, $F+W_\phi=0$. In this case, one can write the Lagrangian density in the form
\begin{align}\label{model}
	\mathcal{L} = \mathcal{L}_b+\mathcal{L}_f,
\end{align}
where
\begin{align}\label{modelb}
	\mathcal{L}_b = \frac{1}{2} \partial_\mu\phi\partial^\mu \phi
	- \frac12 W_\phi^2,
\end{align}
and
\begin{align}\label{modelf}
	\mathcal{L}_f =  \frac{1}{2}\bar{\psi}\,i\gamma^\mu\partial_\mu\psi
	- \frac{1}{2}W_{\phi\phi}\,\bar{\psi}\psi.
\end{align}
In this paper, we plan to study the fermion system given by the Lagrangian density $\mathcal{L}_f$ in the background of kinklike configurations, the solution of the equation of motion considering $\mathcal{L}_b$. Here, we will only focus on the behavior of the Dirac field in the background of the topological structures, leaving the harder issue of the fermion backreaction on the kinklike configurations to be considered elsewhere. Similar investigations dealing with Majorana bound states were carried out before, for instance, in the case of Josephson vortices in topological superconductors \cite{m1} and for a generic two-dimensional field theory that supports kinklike structures \cite{m2}.

The standard bosonic model is described by
\begin{equation}
W=\phi-\frac13 \phi^3,
\end{equation}
and one recalls that we are using dimensionless fields and coordinates in the current work. In this case, the topological structure has the standard kink profile
\begin{equation}
\phi(x)=\tanh(x).
\end{equation}
We will consider two distinct bosonic systems, one described by
\begin{align}\label{modelp}
	V(\phi)=\frac{1}{2}\, p^2\left(\phi^{1-1/p} - \phi^{1+1/p}\right)^2,\;\;\;\;\; W_\phi=p\left(\phi^{1-1/p}-\phi^{1+1/p}\right),
\end{align}
with $p$ being a positive odd integer, and the other by
\begin{align}\label{modeln}
	V(\phi)=\frac{1}{2} \left(1- \phi^{2n} \right)^2, \;\;\;\;\;W_\phi=1-\phi^{2n},
\end{align}
where $n$ is a positive integer. Similar investigations have been carried out in Refs.~\cite{m2,farid,Graham,Dong1,Dong2,AA}.

It is worth mentioning that both potentials \eqref{modelp} and \eqref{modeln} reproduce the $\phi^4$ model in the case $p=n=1$. Moreover, considering the field $\phi$ to be static, the Euler-Lagrange equations of the boson-fermion system are given by 
\begin{align}\label{equations_of_motion_1}
     i\gamma_\mu\partial^\mu \psi -W_{\phi \phi}\,\psi &= 0,\nonumber\\
     \phi'' +  W_{\phi} W_{\phi \phi} &= 0,
\end{align}
where the bosonic solution is background for the fermion field. Defining  $\psi =e^{-i E t} \begin{pmatrix}
		\psi_1  \\
		\psi_2 
	   \end{pmatrix} $, the equations become
\begin{align}\label{equations_of_motion}
     E\ \psi_1 + \psi_2' -W_{\phi \phi}\,\psi_2 &= 0,\nonumber\\
     E\ \psi_2 - \psi_1' -W_{\phi \phi}\, \psi_1 &= 0,\nonumber\\
     \phi'' + W_{\phi} W_{\phi \phi}  &= 0,
\end{align}
after choosing the representation for the Dirac matrices as $\gamma^0 = \sigma_1$, $\gamma^1 = i \sigma_3 $ and $\gamma^5 = \sigma_2$, with the prime denoting differentiation with respect to the $x$ coordinate. 

The two models possess parity symmetry, and as a consequence the upper and lower components of the Dirac field have opposite parities regardless of the value of $p$ and $n$. Besides that, the system has charge and particle conjugation symmetries. The charge conjugation operator is
$\gamma^1$, which is also the particle conjugation operator. This means that the negative and positive energy spectrums are mirror images of each other around the line $E=0$.

Taking $E = 0$, the fermionic equations become
\begin{align}\label{equations_of_motion}
     \psi_2' -W_{\phi \phi}\,\psi_2 &= 0,\nonumber\\
     \psi_1' +W_{\phi \phi}\,\psi_1 &= 0.
\end{align}
It turns out that these two equations can be easily solved as functions of $\phi$, yielding the following zero mode
\begin{align}\label{a15}
     \psi_1(x) &= c_1\,\mbox{e}^{- \int^x\, W_{\phi \phi} \left[\phi(x')\right]\, dx'},\nonumber\\
     \psi_2(x) &= c_2\,\mbox{e}^{\,\int^x \, W_{\phi \phi} \left[\phi(x')\right]\, dx'}.
\end{align}
Defining $g(x) \equiv \mbox{e}^{-\int^x \, W_{\phi \phi} \left[\phi(x')\right]\, dx'}$, either $g(x\to\pm\infty) = 0$ and consequently $g^{-1}(x\to\pm\infty)$ diverges or $g^{-1}(x\to\pm\infty) = 0$ and as a result $g(x\to\pm\infty)$ diverges. Then, to have a normalized zero mode one requires that either $c_1 = 0$ or $c_2 = 0$ which is similar for the two new models as well as the case investigated before in \cite{jr}.

We go on and investigate the threshold or half-bound states, which are states where the fermion field goes to a constant when $x \to \pm \infty$. Although the wave function is finite when $x\to\pm\infty$, these states do not decay fast enough to be square-integrable \cite{Graham,Dong1,Dong2}. Anyway, to find threshold energies we solve the system of equations at $x\to\pm\infty$. We write $\psi$ in the form 
$$\psi(x\to\pm\infty) = e^{-i E t} \begin{pmatrix}
							a_1 \\
							a_2
				   	  \end{pmatrix},$$
where $a_1$ and $a_2$ are arbitrary constants. Under the conditions $\phi(x\to \infty) = 1$ and 
$$\psi'(x\to\infty) = \begin{pmatrix}0 \\0\end{pmatrix},$$
the set of equations of motion become
\begin{align}\label{thresholdgeneral}
     E_{\mathrm {th}}\,a_1  - W_{\phi \phi}\left(\phi=1\right)\,\,a_2 &= 0,\nonumber\\
     E_{\mathrm {th}}\,a_2  - W_{\phi \phi}\left(\phi=1\right)\,\,a_1 &= 0,
\end{align}
resulting in $E_{\mathrm {th}}=\pm W_{\phi \phi}\left(\phi=1\right)$ as threshold energies. In the model \eqref{modelp}, the threshold energies are $E_{\rm th}=\pm2$; they do not depend on $p$ and reproduce the values of the standard $\phi^4$ model. In the model \eqref{modeln} they are $E_{\rm th}=\pm2n$ and depend linearly on $n$.

\section{New models}
\label{sec:models}

Let us now focus on the new models that appear in \cite{bmm,blmm}, described by \eqref{modelp} and \eqref{modeln}. We first deal with the scalar model that supports the two-kink solution, and then with the case that leads to the compact kinklike configuration.

\subsection{The two-kink background}

\begin{figure}[h]
\begin{center}
	{\includegraphics[width=0.5\textwidth]{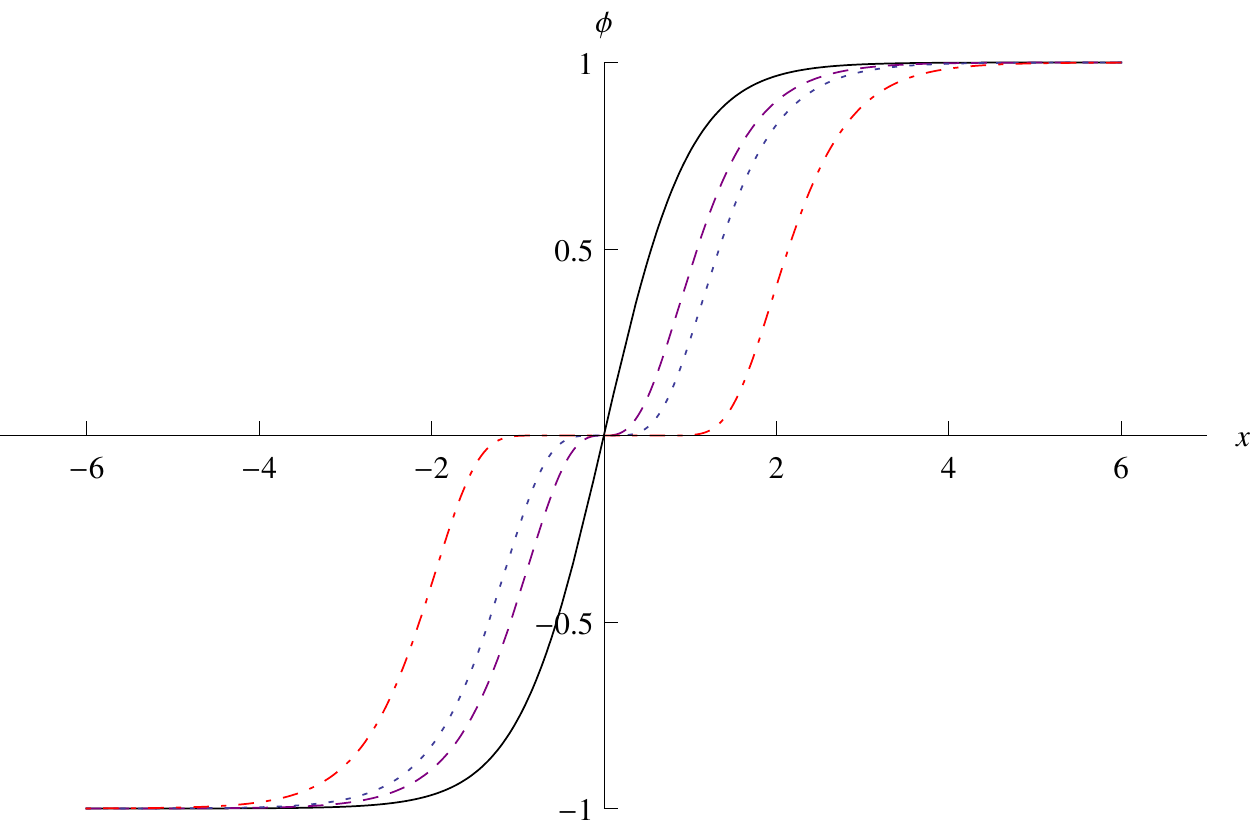}}
	\caption{The topological structure $\phi(x)$ that appears in the model \eqref{modelp}, displayed by solid, dashed, dotted and dot-dashed lines for $p=1,3,5$, and $25$, respectively.}
    \label{fig1}
\end{center}
\end{figure}

Let us consider the Dirac field interacting with the kinklike structure that appears in the model \eqref{modelp}. The two-kink solution has the form $\phi(x)=\tanh^p(x)$. In Fig.~\ref{fig1} we display the solution for the cases $p=1,3,5$ and $25$.

As one can see, the greater the value of $p$, the wider the step at the center of the structure, around the origin $x=0$; it resembles two kinks joining at the center, so it is sometimes called a two-kink configuration.

\begin{figure}[h]   
  \centering
{\includegraphics[width=0.5\textwidth]{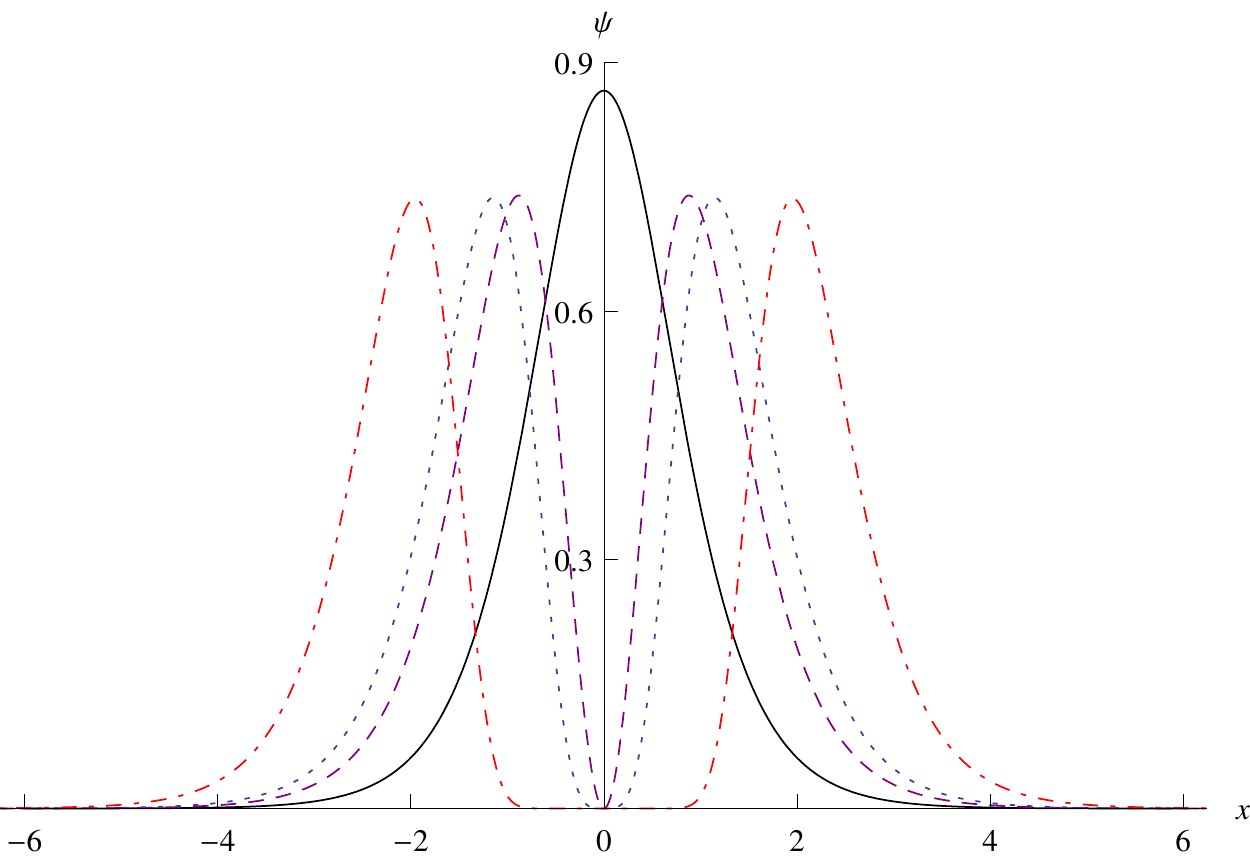}}
  \caption{The fermionic zero mode displayed by solid, dashed, dotted and dot-dashed lines for $p=1,3,5,$ and $25$, respectively.}
  \label{fig2}
\end{figure}

The equation of motion for the fermion field has the form
\begin{align}\label{equations_of_motion_1}
     i\gamma_\mu\partial^\mu \psi -p  \left[(1-1/p)\phi^{-1/p} - (1+1/p)\phi^{1/p}\right]\,\psi &= 0.
\end{align}
where $\phi=\tanh^p(x)$. Defining  $\psi =e^{-i E t} \begin{pmatrix}
		\psi_1  \\
		\psi_2 
	   \end{pmatrix} $, the equation becomes
\begin{align}\label{equations_of_motion}
     E\ \psi_1 + \psi_2' -p  \left[(1-1/p)\phi^{-1/p} - (1+1/p)\phi^{1/p}\right]\,\psi_2 &= 0,\nonumber\\
     E\ \psi_2 - \psi_1' -p  \left[(1-1/p)\phi^{-1/p} - (1+1/p)\phi^{1/p}\right]\,\psi_1 &= 0,
\end{align}
in terms of the two components. These coupled equations are not analytically solvable for arbitrary $p$, except for the zero mode and threshold energies as we discussed before. For the zero energy we have
\begin{align}\label{a15}
     \psi_1(x) &= c_1\,\mbox{e}^{-p \int [(1-1/p)\phi^{-1/p}(x') - (1+1/p)\phi^{1/p}(x')]\, dx'},\nonumber\\
     \psi_2(x) &= c_2\,\mbox{e}^{\,p\int [(1-1/p)\phi^{-1/p} (x')- (1+1/p)\phi^{1/p}(x')]\, dx'}.
\end{align}
For arbitrary $p$, the normalized zero mode is given by
\begin{align}\label{a18}
	\psi(x) = \sqrt{4p^2-1}
			\begin{pmatrix}
				0 \\
				\text{tanh}^p\left(x\right)\text{csch}\left(2x\right)
		  	\end{pmatrix}.
\end{align}
In the case $p=1$, the normalized zero mode is
\begin{align}\label{a18}
	\psi(x) = \sqrt{\frac{3}{4}}
			\begin{pmatrix}
				0 \\
				\text{sech}^{2}\left(x\right)
		  	\end{pmatrix},
\end{align}
which exactly matches the analytical solution in the standard kink background; see, e.g., Ref.~\cite{farid}. 
In Fig.~\ref{fig2} we display the zero mode for each of the cases $p=1,3,5,$ and $p=25$. One sees that for $p=3,5,...$ the zero mode seems to split into two similar portions around the origin. As $p$ grows, the flat region around the center of the solution and as a consequence the distance between the portions of the zero mode increases. We will show shortly, this is not the case when one considers the Yukawa coupling.

As shown below Eq.~\eqref{thresholdgeneral}, in this case the threshold energies are $E_{\mathrm {th}}=\pm2$. We then go on to search for other bound states, but this requires a numerical procedure. Here we use the Runge-Kutta-Fehlberg method of order $5$. Our numerical calculations show that for $p>1$ there is no fermionic bound state except the zero mode. However, for $p=1$ the system has three bound states matching the analytic solutions $E=0,\pm\sqrt3$, within the numerical precision that we are considering in this work. Moreover, we have noted that the stability equation for the bosonic solution is
\begin{align}
\left(-\frac{d^2}{dx^2}+\left.\frac{d^2 V(\phi)}{d\phi^2}\right|_{\phi_s(x)}\right) \eta_n(x)=\omega_n^2  \eta_n(x),
\end{align}
and it gives the same spectrum as the one for the fermionic bound states, as expected, within our numerical limitations.

\subsection{The compact kinklike background}

Let us now consider the model with potential \eqref{modeln}, as originally introduced in \cite{blmm}.
In order to guarantee a well-defined energy for the kinklike configuration, we have to consider $\phi(x\to\pm\infty)=\pm1$.
The equations of motion for the system with a static bosonic structure can be written as
\begin{align}\label{eomn}
     E\ \psi_1 + \psi_2' +2n \,\phi^{2n-1}\,\psi_2 &= 0,\nonumber\\
     E\ \psi_2 - \psi_1' +2n \,\phi^{2n-1}\,\psi_1 &= 0,\nonumber\\
    \phi'' + 2n\phi^{2n-1}\left(1-\phi^{2n}\right)  &= 0,
\end{align}
choosing the same representation for the Dirac matrices. 

\begin{figure}[h]
\begin{center}
{\includegraphics[width=0.5\textwidth]{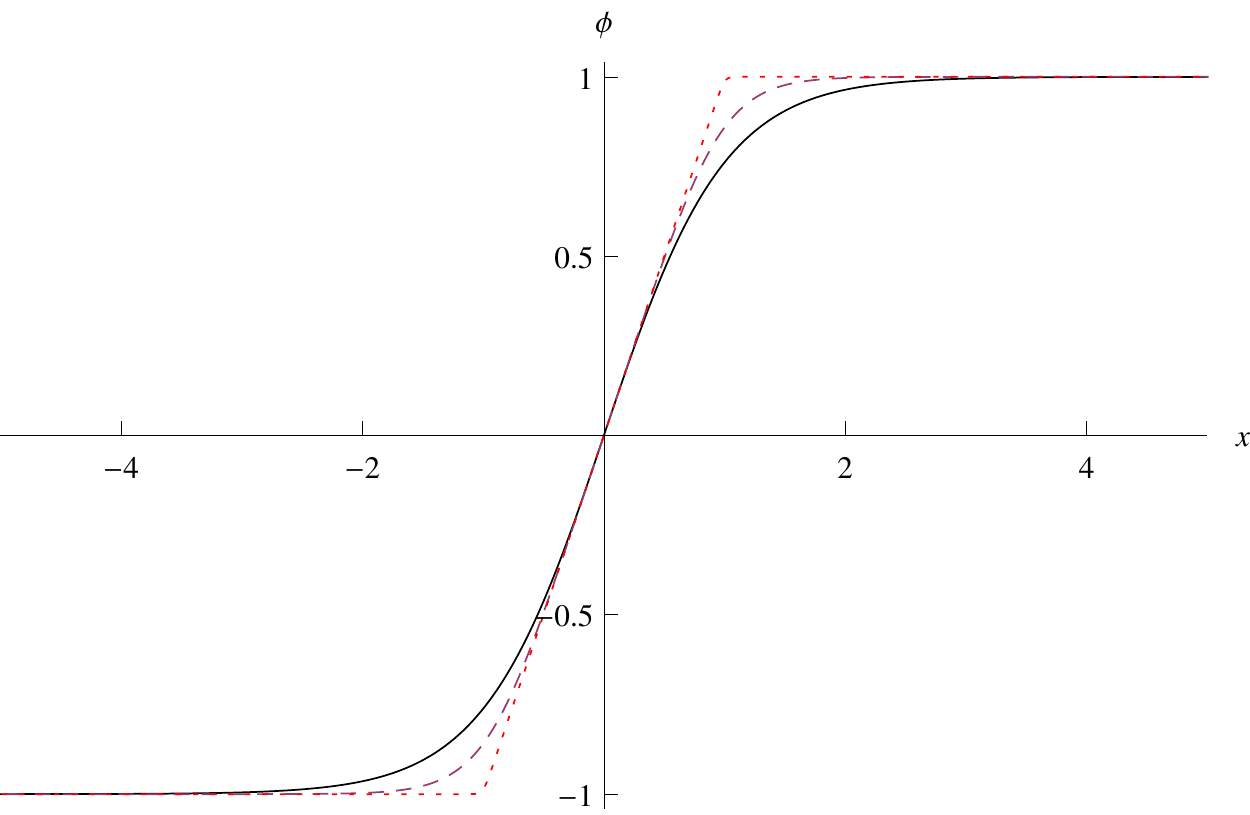}}
	\caption{The topological structure $\phi(x)$ that appears in the model \eqref{modeln}, displayed by solid, dashed and dotted lines for $n=1,2,$ and $25$, respectively.}
    \label{fig3}
\end{center}
\end{figure}

In the case $n=1$, the kinklike solution for the bosonic equation is $\phi(x)= \tanh(x)$.
However, in the case of $n=2,3,...$ one has to find the solution numerically \cite{blmm}. To illustrate how the bosonic configuration behaves, in Fig.~\ref{fig3} we show the numerical solution for $n=1,2,$ and $25$. As it shows very clearly, the solution shrinks to a compact interval in the $x$ coordinate as $n$ increases to larger and larger values.

\begin{figure}[h]
\begin{center}
{\includegraphics[width=0.5\textwidth]{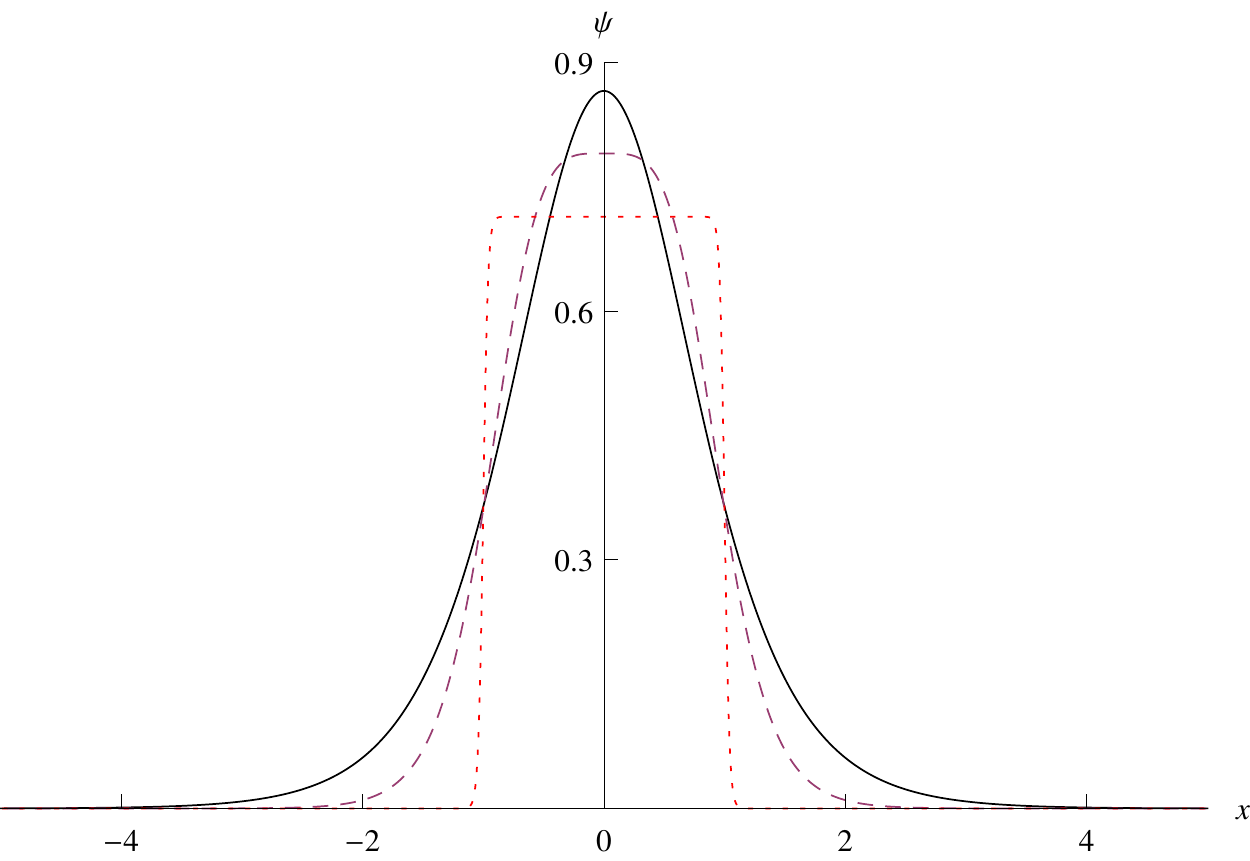}}
	\caption{The fermionic zero mode, displayed by solid, dashed and dotted lines for $n=1,2,$ and $25$, respectively.}
    \label{fig4}
\end{center}
\end{figure}

Let us now investigate how the fermion field behaves under such background. We first consider the zero energy states, taking $E = 0$ in Eqs.~\eqref{eomn}. The solutions are 
\begin{align}\label{a15}
     \psi_1(x) &= c_1\,\mbox{e}^{2 n \int\, \phi^{2n-1} (x')\, dx'},\nonumber\\
     \psi_2(x) &= c_2\,\mbox{e}^{-2 n \int \phi^{2n-1} (x')\, dx'}.
\end{align}
For $n=1$, one can solve the above equations analytically \cite{AA}. In this case the normalized wave function is
\begin{align}\label{a18}
	\psi(x) = \sqrt{\frac{3}{4}}
			\begin{pmatrix}
				0 \\
				\text{sech}^{2}\left(x\right)
		  	\end{pmatrix}.
\end{align}
For arbitrary $n$ we have to find the zero mode numerically. We show in Fig.~\ref{fig4} the fermion zero mode for $n=1,2,$ and $n=25$, in the background of the solutions shown in Fig.~\ref{fig3}. One can notice that, as the topological structure shrinks by increasing $n$, the fermionic zero mode also shrinks into the same compact interval. 

\begin{figure}[h]
  \centering
\includegraphics[width=0.5\textwidth]{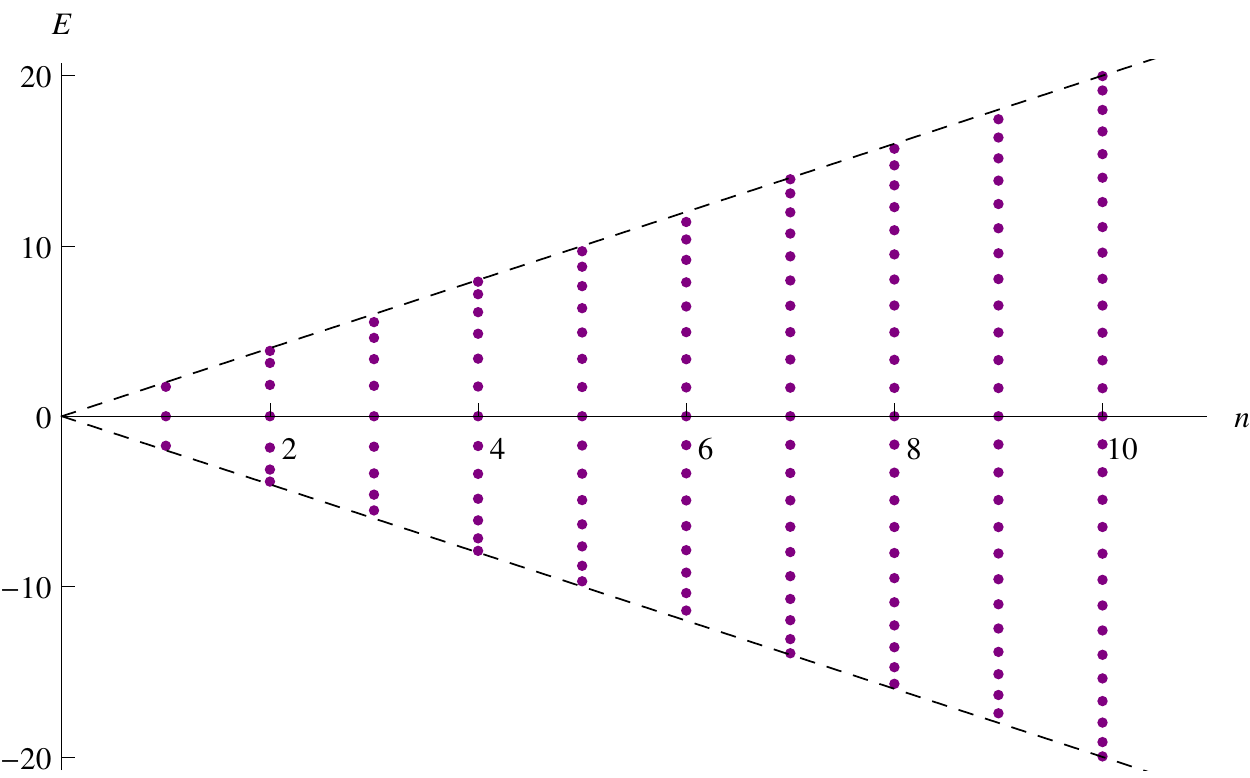} 
\caption{The fermionic bound energies as a function of $n$. The dashed lines show the threshold energies.}
  \label{fig5}
\end{figure}

In order to find the other bound states and bound energies for arbitrary $n$, we need to adopt a numerical method. We use the same Runge-Kutta-Fehlberg method of order 5 that we used before. The energy spectrum as a function of $n$ is shown in Fig.~\ref{fig5}. In this figure we also show the threshold energy lines, which are $E_{\rm th}=\pm2n$, and one notices that all the bound energies are confined in between the two threshold energies; outside this region there are scattering states. It is interesting to note that the bound energy spectrum is symmetric around the $E=0$ line, reflecting the fact that the system has particle conjugation symmetry. Solving the stability equation for the corresponding bosonic portion gives the same spectra as the ones for the fermionic bound states, within the numerical limitation, as expected.

\begin{figure}[h]
  \centering
\includegraphics[width=0.5\textwidth]{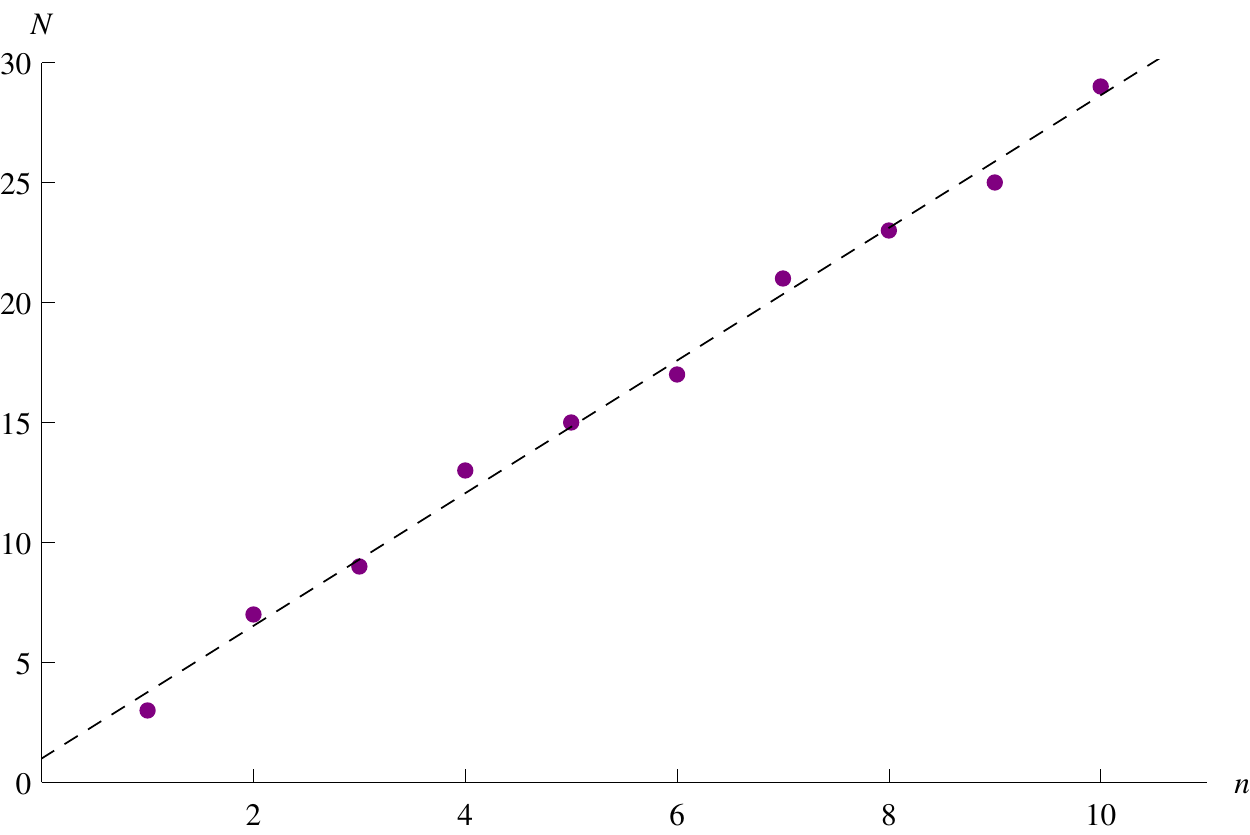} 
\caption{The number of bound states $N$ as a function of $n$. The dashed line shows the linear fit.}
  \label{fig6}
\end{figure}

For $n=1$, the system is analytically solvable and it is easy to show that there are three fermionic bound states with energies $E=0$ and $E=\pm\sqrt{3}\,$ and two threshold ones with energies $E=\pm2\,$ (see \cite{farid,AA}). Besides analytical results $E=0$ and $E=\pm2$, our numerical method results in the bound energies $E=\pm 1.7316$. As can be seen, our results are the same as the analytical ones, within the numerical precision, which is a consistency check for the system we are studying here. It is interesting to note that as the integer $n$ increases, the number of bound states also increases. In Fig.~\ref{fig6} we show the number of bound states $N$ as a function of $n$. The dashed line is the fit to the points in this figure which is equal to $1 + 2.76364 \,n$. The number of bound states for each $n$ is exactly equal to the number of bound states of the stability potential related to the bosonic potential, as displayed in Fig. 5 of the work \cite{blmm}. We remark here that in Ref.~\cite{blmm} the authors show the result for $\omega_n^2$. In the current work, however, we count both the positive and negative bound states as separate fermionic bound states.

\begin{figure}[h]
  \centering
  \begin{tabular}{cc}
      \begin{subfigure}[b]{0.35\textwidth}
		\includegraphics[width=\textwidth]{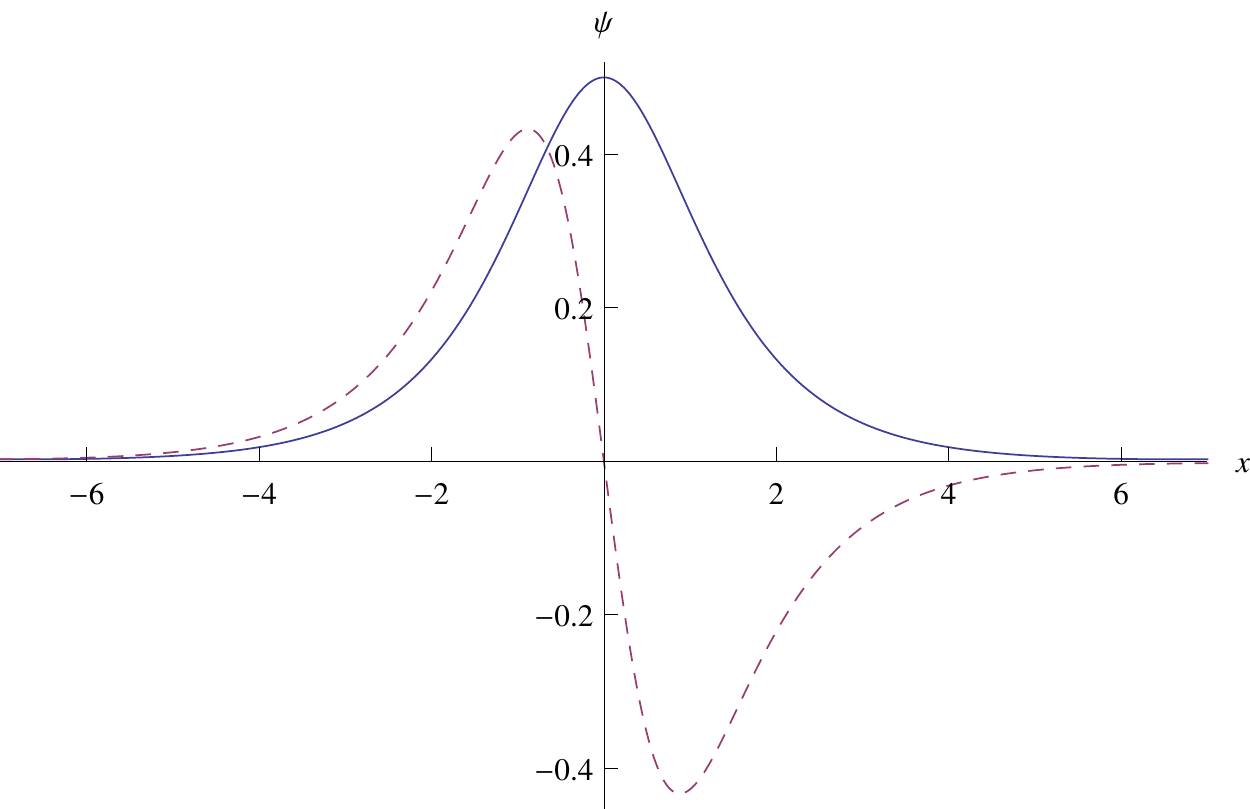}
        \caption{The components $\psi_1$ and $\psi_2$ of the massive bound state for $n=1$, at the bound energy $E=1.7316$.}
        \label{fig:wavefuncn1}
    \end{subfigure}
 &        \begin{subfigure}[b]{0.35\textwidth}
		\includegraphics[width=\textwidth]{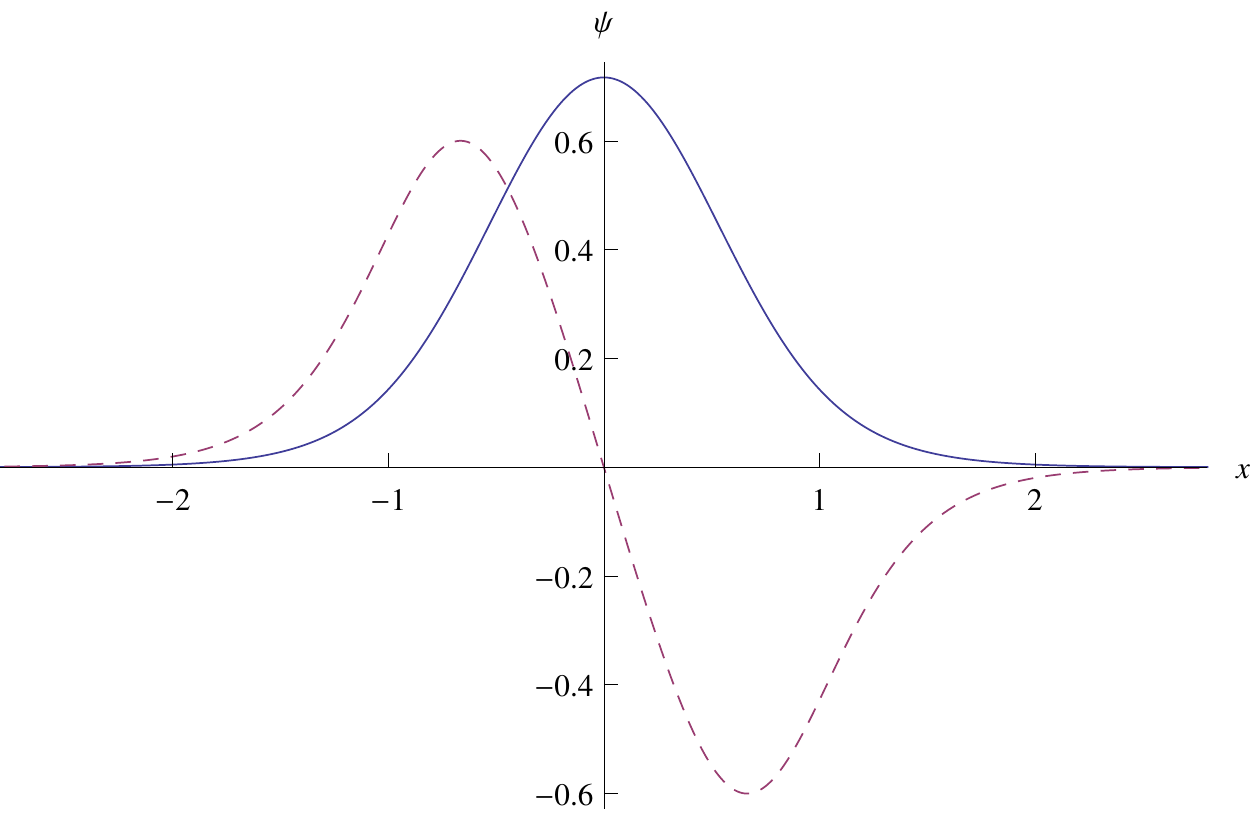}
        \caption{The components $\psi_1$ and $\psi_2$ of the massive bound state for $n=2$, at the bound energy $E=1.8409$.}
        \label{fig:wavefuncn21}
    \end{subfigure}\\
      \begin{subfigure}[b]{0.35\textwidth}
		\includegraphics[width=\textwidth]{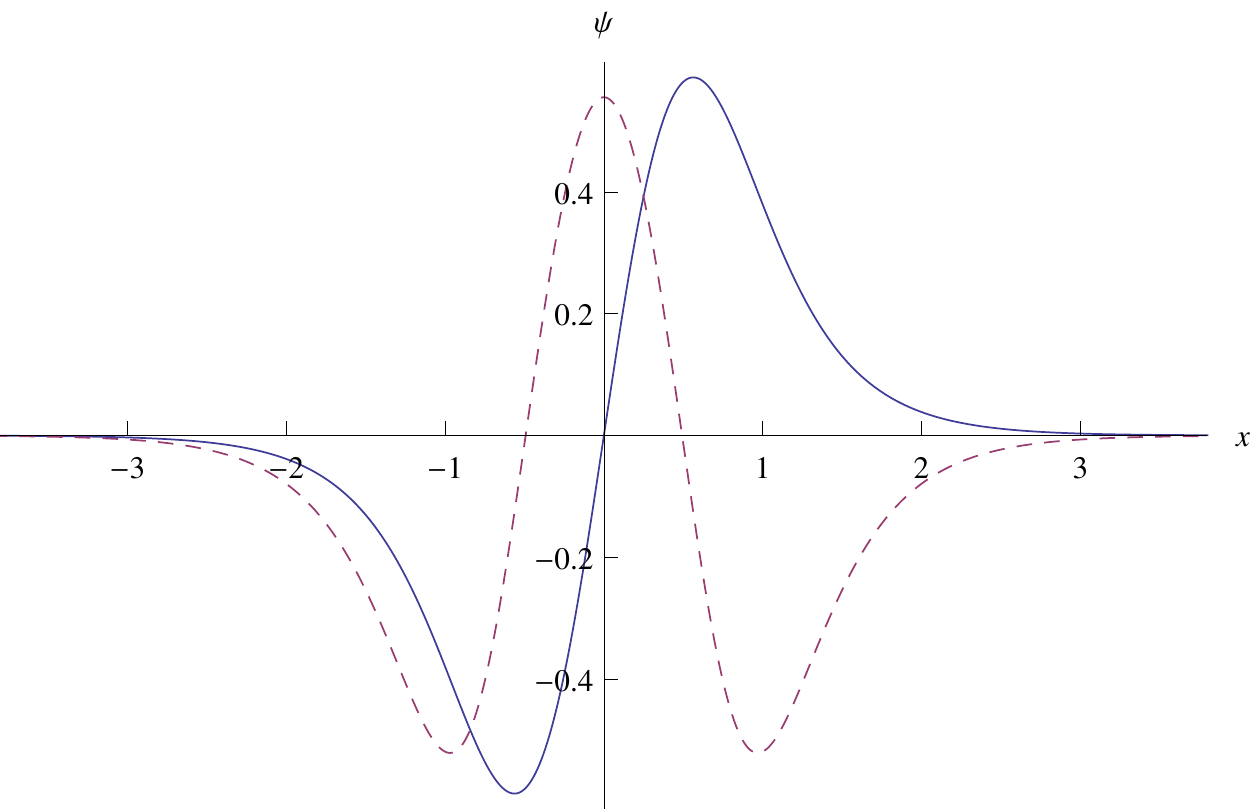}
        \caption{The components $\psi_1$ and $\psi_2$ of the massive bound state for $n=2$, at the bound energy $E=3.1298$.}
        \label{fig:wavefuncn22}
    \end{subfigure} &   
    \begin{subfigure}[b]{0.35\textwidth}
		\includegraphics[width=\textwidth]{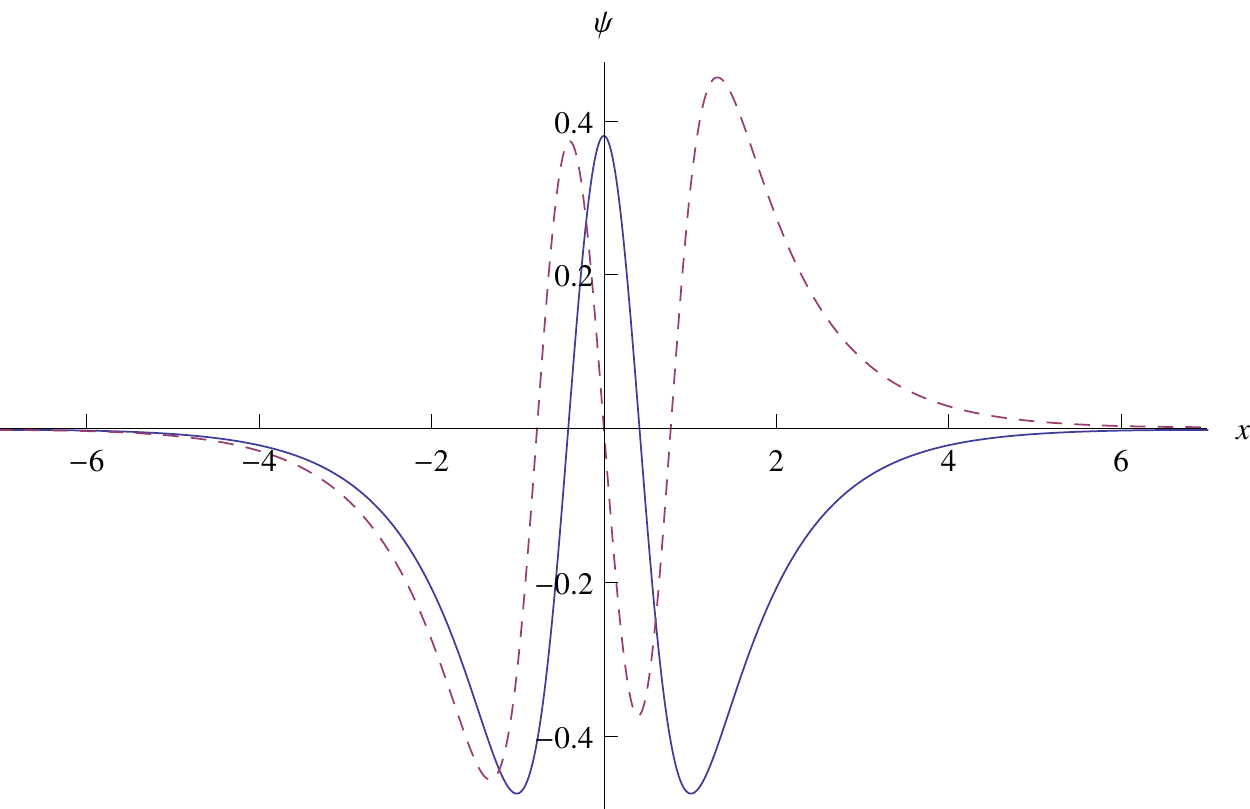}
        \caption{The components $\psi_1$ and $\psi_2$ of the massive bound state for $n=2$, at the bound energy $E=3.8359$.}
        \label{fig:wavefuncn23}
    \end{subfigure}
  \end{tabular}
  \caption{The two components of the massive fermionic bound states, displayed for $n=1$ and $2$, at the corresponding bound energies.}
  \label{fig7}
\end{figure}

We notice that in this model, the number of massive bound states increases linearly with $n$. This behavior does not appear in the previous model, since it does not support massive bound states for any $p=3,5,...$\,. In order to further study the massive fermionic bound states, let us now investigate the cases $n=1$ and $n=2$, for example. In Fig.~\ref{fig:wavefuncn1} one shows the two components of the wave function, $\psi_1$ and $\psi_2$, for $n=1$ at the bound energy $E=1.7316$ which matches the analytical solution
\begin{align}
\psi(x) =  \begin{pmatrix}
\frac{1}{2\cosh(x)} \\
-\frac{\sqrt{3}\tanh(x)}{2 \cosh(x)}
\end{pmatrix}
\end{align}
with $E=\sqrt{3}$, within our numerical precision. Moreover, in Figs.~\ref{fig:wavefuncn21}, \ref{fig:wavefuncn22} and \ref{fig:wavefuncn23} we show the two components of the wave function, for $n=2$ at the bound energies $E=1.8409$, $E=3.1298$ and $E=3.8359$, respectively. For each $n$, the number of the nodes in the wave function increases when the energy increases. This is confirmed by the results displayed in Figs.~\ref{fig:wavefuncn21}, \ref{fig:wavefuncn22} and \ref{fig:wavefuncn23}, for $n=2$. Also, it is clear from the figures that the upper and lower components have opposite parities. 

\subsection{Yukawa coupling}

Let us now recall the experiment implemented in Ref.~\cite{prl}, where current pulses are injected into the magnetic strip that supports the domain wall. If one thinks of electrons interacting with the magnetic arrangement that build the topological structure, one has to model the boson-fermion interaction term to be added to the system. An interesting alternative to the coupling used in the previous section is to consider the model \eqref{model} with ${\cal L}_b$ given by \eqref{modelb}, but with the fermionic portion changed to
\begin{align}\label{modelf2}
	\mathcal{L}_f =  \frac{1}{2}\bar{\psi}\,i\gamma^\mu\partial_\mu\psi
	- \phi\,\bar{\psi}\psi.
\end{align}
with the boson-fermion interaction now controlled by the standard Yukawa coupling. 

Let us now investigate this new possibility. Here the fermionic equations of motion becomes
\begin{align}\label{fermion2k}
     E\ \psi_1 + \psi_2' +2 \,\phi\,\psi_2 &= 0,\nonumber\\
     E\ \psi_2 - \psi_1' +2\,\phi\,\psi_1 &= 0,
\end{align}
where $\phi$ is given analytically by $\phi(x)=\tanh^p(x)$ for the two-kink model, with $p=1,3,5,...$, or numerically in the case of the compact kinklike model. The result is the same as the previous one with supersymmetric coupling when $n=p=1$, which can be used as a consistency check. Moreover, it can be used to study how the results diverge from each other considering these two distinct types of couplings while increasing $n$ and $p$.

For the threshold or half-bound states the set of equations of motion are
\begin{align}\label{thresholdgeneral1}
     E_{\mathrm {th}}\,a_1  + 2\,a_2 &= 0,\nonumber\\
     E_{\mathrm {th}}\,a_2  + 2\,a_1 &= 0.
\end{align}
where $a_1$ and $a_2$ represent the values of the fermionic fields at infinity. This results in $E_{\mathrm {th}}=\pm 2$ for both potentials.

\begin{figure}[h]
\begin{center}
	\includegraphics[width=0.5\textwidth]{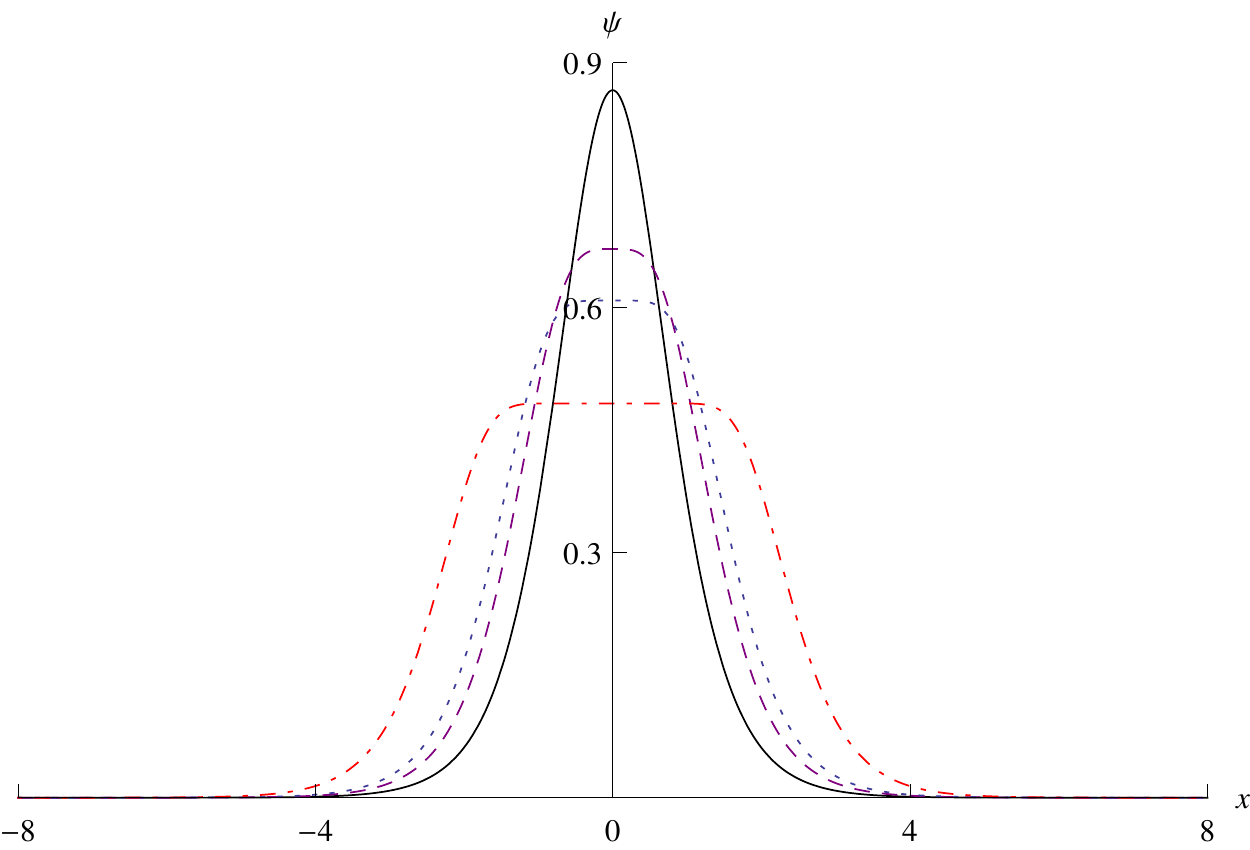}
	\caption{The fermion zero mode for the Yukawa coupling with the two-kink model, displayed by solid, dashed, dotted and dot-dashed lines for the cases $p=1,3,5$, and $p=25$, respectively.}
    \label{fig8}
\end{center}
\end{figure}

The zero mode has to be constructed from 
\begin{align}\label{a15}
     \psi_1(x) &= c_1\,\mbox{e}^{- 2\int^x\, \phi(x')\, dx'},\nonumber\\
     \psi_2(x) &= c_2\,\mbox{e}^{2\int^x \, \phi(x')\, dx'}.
\end{align}
As in the previous cases, one has to choose $c_1=0$ or $c_2=0$ to make the zero mode normalizable. In the two-kink background the fermionic zero mode can be obtained analytically. It is given by
\begin{align}\label{a18}
	\psi(x) = c_p
			\begin{pmatrix}
				0 \\
				\exp\Bigl[-2\,\text{tanh}^{1+p}\left(x\right)\, _2F_1\left(1,\frac{1+p}{2};1+\frac{1+p}{2};\text{tanh}^{2}\left(x\right)\right)\Bigr]
		  	\end{pmatrix}.
\end{align}
where $c_p$ is the normalization factor to be calculated numerically for arbitrary $p$. We show the normalized zero mode in
Fig.~\ref{fig8} for the four specific values $p=1,3,5$, and $p=25$. 
\begin{figure}[h]
\begin{center}
	\includegraphics[width=0.5\textwidth]{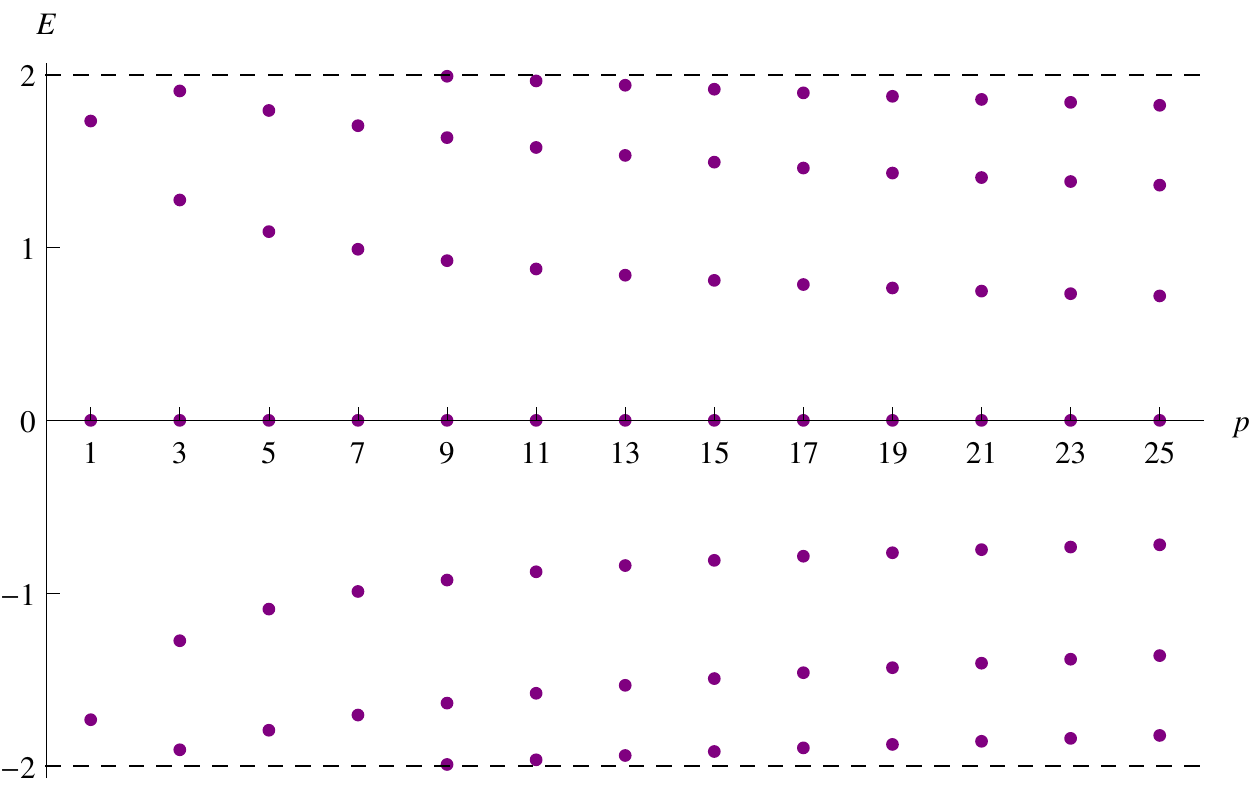}
	\caption{The fermion energy spectrum for the Yukawa coupling with the two-kink model. The dashed lines show the threshold energies.}
    \label{fig9}
\end{center}
\end{figure}
We follow the same numerical procedure, as in the previous case, to find the fermionic bound energy spectrum. The result is shown in Fig.~\ref{fig9}. Compared to the previous case, the number of bound states increases with $p$, although the threshold energies remain the same. In Fig.~\ref{fig10} one quantifies the number of bound states for each $p$.

\begin{figure}[h]
\begin{center}
	\includegraphics[width=0.5\textwidth]{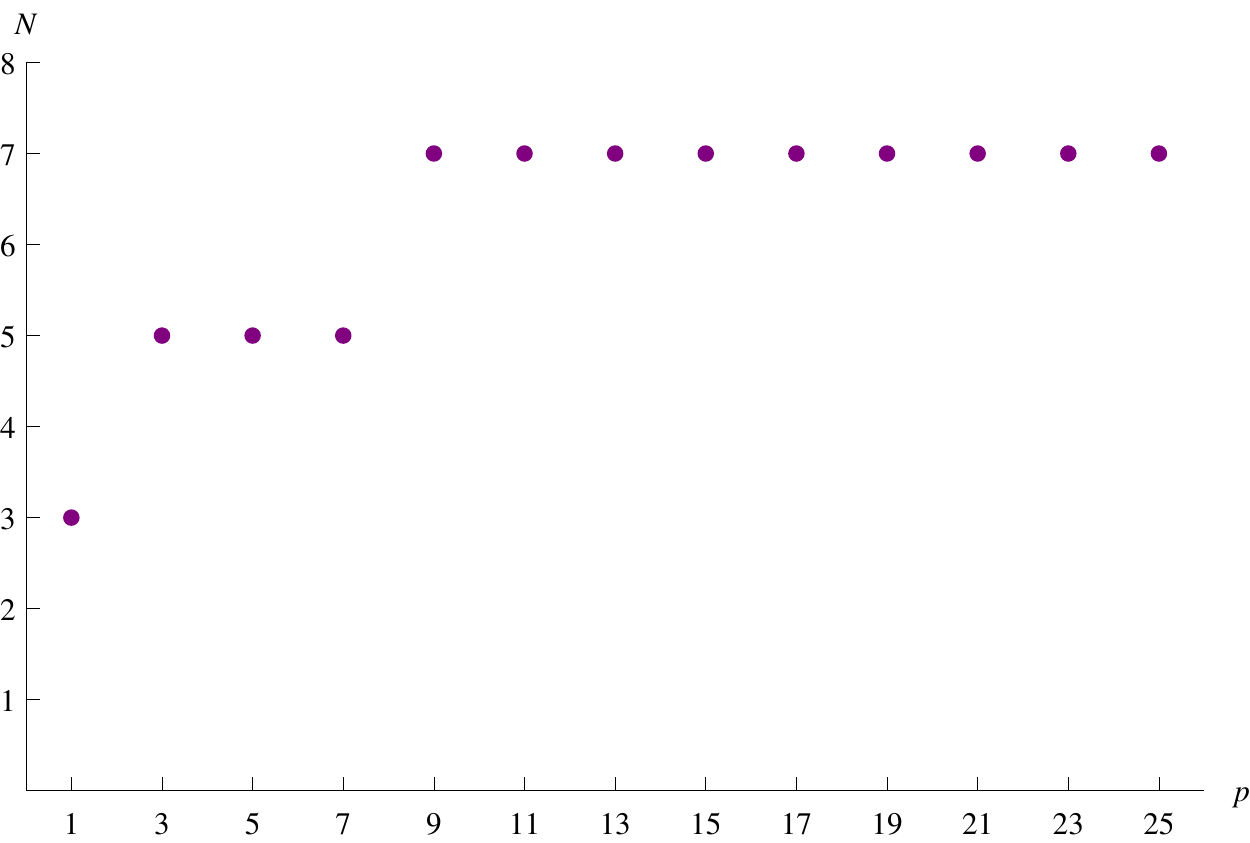}
	\caption{The number of bound states $N$ for the Yukawa coupling with the two-kink model, for several values of $p$.}
    \label{fig10}
\end{center}
\end{figure}
\begin{figure}[h]
\begin{center}
	\includegraphics[width=0.5\textwidth]{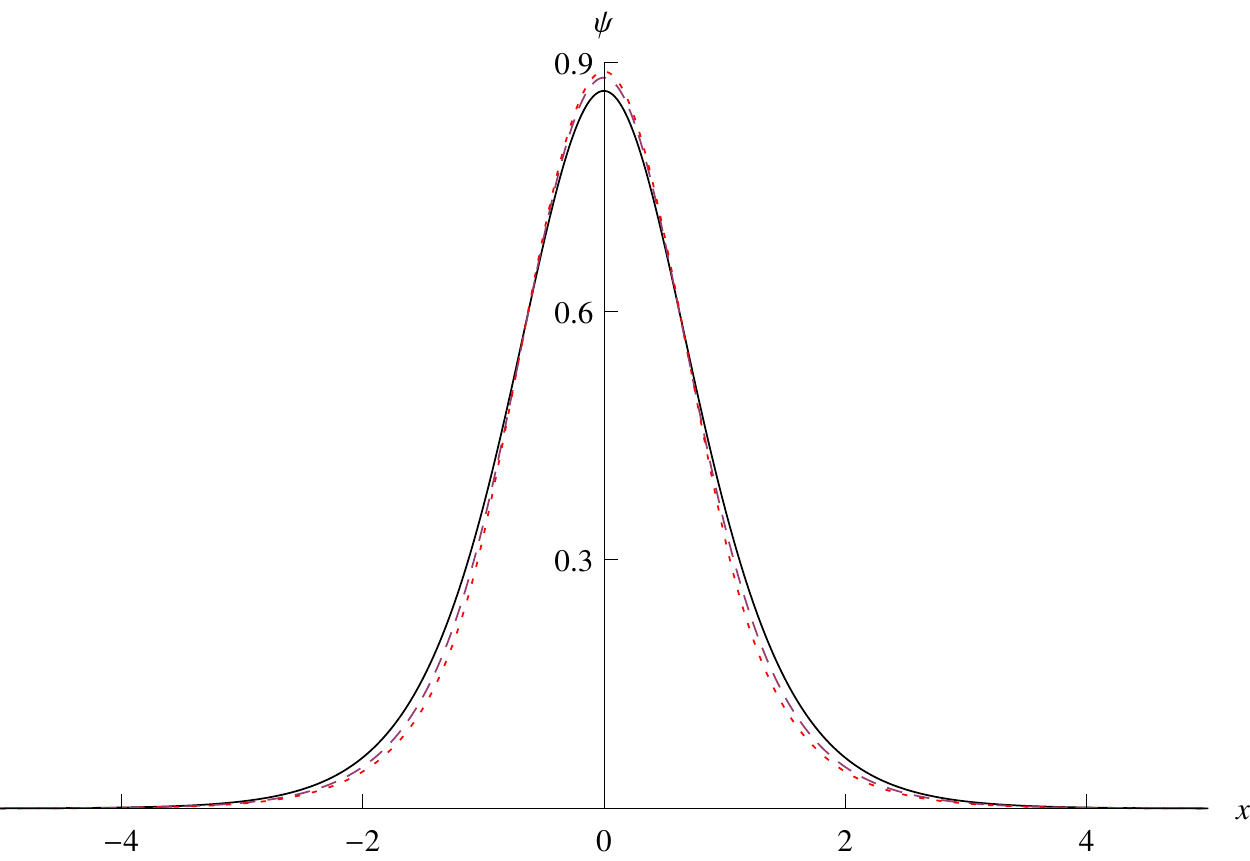}
	\caption{The fermionic zero modes for the Yukawa coupling with the compact kinklike model, displayed by solid, dashed and dotted lines, for $n=1$, $n=2$ and $n=25$, respectively.}
    \label{fig11}
\end{center}
\end{figure}

In the case of the compact kinklike background we have to implement numerical calculations. The zero mode is now shown in
Fig.~\ref{fig11}. Here we see that the effect of the compact kinklike structure considering the Yukawa coupling is not as bold as before.
\begin{figure}[h]
\begin{center}
	\includegraphics[width=0.5\textwidth]{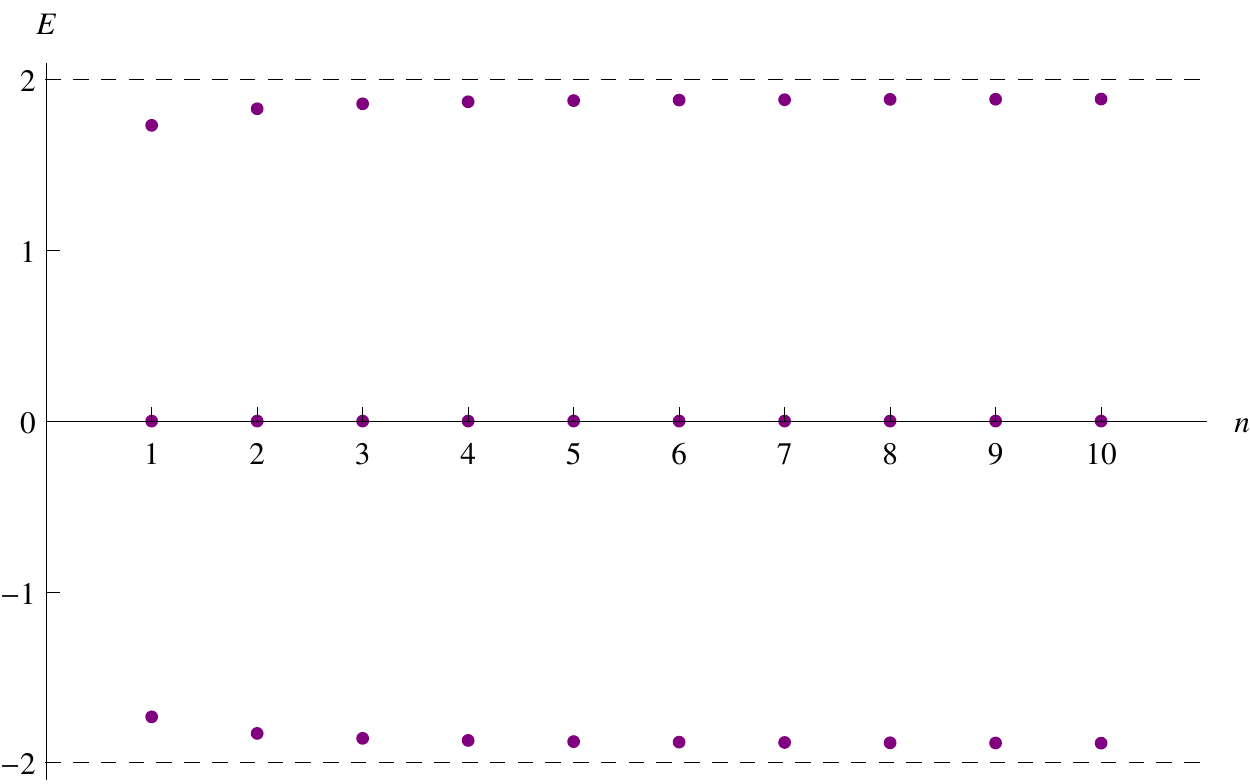}
	\caption{The fermion energy spectrum for the Yukawa coupling with compact kinklike model. The dashed lines show the threshold energies.}
    \label{fig12}
\end{center}
\end{figure}
To find the other bound energies and states we use the same numerical method. In Fig.~\ref{fig12} one shows the fermionic bound energies for several values of $n$, with the solid lines showing the threshold energies. It is interesting to note that the number of the bound states does not depend on $n$; the spectrum is composed of three bound states, independently of the values of $n$. This is completely different from the behavior found in the previous case, with the coupling suggested by supersymmetry.
\section{Comments and Conclusions}
\label{sec:end}

In this work we investigated how the bound states of fermions behave as we change the topological background which is described by the bosonic field $\phi$ in the two-dimensional spacetime. We considered two distinct backgrounds, one described by the two-kink structure that appears in the model with potential \eqref{modelp}, investigated in \cite{bmm}, and the other described by the potential \eqref{modeln}, that supports a compact kinklike structure \cite{blmm} which is driven by a positive integer $n$ controling the power of the self-interaction in the bosonic sector of the system.

It is important to mention that although the new background configurations do not change the topology, they engender distinct profiles: in the first case, the two-kink solution has the same asymptotic behavior, but is very different from the standard kink at its center. The result when the coupling is suggested by supersymmetry is that the modification at the center of the two-kink background works to eliminate massive bound states for the fermion field. Thus, although the standard and the two-kink structure are able to support the same fermionic zero mode for the case $p=1$, the two-kink background is not able to support massive fermionic bound states for larger $p$. In the second case, however, for the compact kinklike background the fermion field is capable of supporting a diversity of massive bound states, the quantity being proportional to $n$, so it increases as $n$ increases to larger and larger values.

We studied other possibilities, with the boson-fermion interaction being controlled by the Yukawa coupling. Here we noted that, when the topological structure is the two-kink, the fermion field presents the zero mode and several other massive bound states, the number depending on the value of the parameter $p$. However, if the topological structure is of the compact kinklike form, the number of bound states is always three, the zero mode and two other massive bound states, irrespective of the value of $n$. 

The current study shows that modifications on the geometric form of the topological structures change the fermionic behavior, adding or removing massive bound states from the spectrum of excitations, and this may play a role in the construction of electronic devices at the nanometric scale. Further investigations are required to add more information on the basic properties concerning the boson-fermion interaction in two-dimensional systems similar to the ones studied in this work. Also, the behavior of fermions on planar topological objects such as vortices and skyrmions seems to be of current interest. We hope to report on these and other related issues in the near future.

\section*{Acknowledgments}
DB would like to thank the Brazilian agency CNPq for partial financial support, under the contracts 455931/2014-3 and 306614/2014-6. AM acknowledges PNPD/CAPES for the financial support. 


\end{document}